# Surface doping of rubrene single crystals by molecular electron donors and acceptors

Christos Gatsios[1], Andreas Opitz[1,*], Dominique Lungwitz[1], Ahmed E. Mansour[2], Thorsten Schultz[2], Dongguen Shin[1], Sebastian Hammer[3,4], Jens Pflaum[3,5], Yadong Zhang[6], Stephen Barlow[6], Seth R. Marder[6,7] and Norbert Koch[1,2,*]

[1]Institut für Physik & IRIS Adlershof, Humboldt-Universität zu Berlin, 12489 Berlin, Germany

[2]Helmholtz-Zentrum Berlin für Materialien und Energie GmbH, 12489 Berlin, Germany

[3]Experimentelle Physik VI, Julius-Maximilians-Universität Würzburg, 97074 Würzburg, Germany

[4]Center for the Physics of Materials, Departments of Physics and Chemistry, McGill University, Montreal, Qc, Canada

[5]Center for Applied Energy Research e.V., Magdalene-Schoch-Str. 3, 97074 Würzburg, Germany

[6]Renewable and Sustainable Energy Institute (RASEI), University of Colorado Boulder, Boulder, CO 80309, USA

[7]Department of Chemical and Biological Engineering and Department of Chemistry, University of Colorado Boulder, Boulder, CO 80309, USA

*E-Mail: norbert.koch@physik.hu-berlin.de, andreas.opitz@hu-berlin.de

## Abstract

The surface molecular doping of organic semiconductors can play an important role in the development of organic electronic or optoelectronic devices. Single-crystal rubrene remains a leading molecular candidate for applications in electronics due to its high hole mobility. In parallel, intensive research into the fabrication of flexible organic electronics requires the careful design of functional interfaces to enable optimal device characteristics. To this end, the present work seeks to understand the effect of surface molecular doping on the electronic band structure of rubrene single crystals. Our angle-resolved photoemission measurements reveal that the Fermi level moves in the band gap of rubrene depending on the direction of surface electron-transfer reactions with the molecular dopants, yet the valence band dispersion remains essentially unperturbed. This indicates that surface electron-transfer doping of a molecular single crystal can effectively modify the near-surface charge density, while retaining good charge-carrier mobility.

## Introduction

Electrical doping in organic semiconductors involves the transfer of electrons between a dopant and the host semiconductor for generating free charge carriers within the host material. The addition of electrons or holes to the semiconductor is accompanied by an energy shift of the Fermi level (electrochemical potential of electrons) relative to the valence and conduction band edges. Manipulating the free charge-carrier concentration, and thus the Fermi level, presents a key strategy for modifying and enhancing the electrical properties of organic semiconductors. Previous efforts employing atomic dopants such as halogens or alkali metals resulted mainly in unstable devices due to the high diffusivity, chemical reactivity, and poor air stability of such dopants [1–5]. In contrast, molecular dopants can be superior doping agents for organic semiconductors due to a number of factors. For instance, their larger size compared to atomic dopants reduces the Coulombic interactions with the contributed charge carriers and the chance of diffusion into the semiconductor's bulk. Also, their moderate sublimation temperature is beneficial for processing and device integrity during fabrication [6–11].

In parallel with the development of tailored, more efficient dopants, the field of organic semiconductors has advanced noticeably. New classes of thienoacene and pyrene-based organic semiconductors with field-effect charge-carrier mobilities as high as 30 $cm^2V^{-1}s^{-1}$ were introduced, paving the way for sustainable flexible electronic devices [12–14]. However, the ideal performance of organic electronic devices is severely impacted by high contact resistances and charge-carrier trapping. Several studies indicated that doping can effectively circumvent both obstacles. For example, adding a thin layer of dopants between the metal contact and the semiconductor can effectively reduce the charge injection barriers [15–17]. Passivation of trap states, which can improve the transport characteristics by lowering the threshold voltage, is a further benefit of doping [18–21]. Also, low to moderate doping of the channel can be beneficial as long as the conductivity does not increase at the cost of a decreased current on/off ratio [22]. A previous work by Lüssem *et al*. demonstrated successful doping of pentacene thin films by evaporating thin layers of n-type and p-type dopants, making depletion and inversion operation modes possible [23]. Other works have also reported the change in the field-effect transistor polarity after doping with molecular electron donors and acceptors [24–26].

Here, we investigate the surface doping of rubrene single crystals, a benchmark organic semiconductor. As illustrated in Figures 1a and 1b, rubrene has a distinct chemical structure and an orthorhombic single-crystalline structure (space group Cmca) [27]. Notably, these characteristics facilitate a high hole mobility along the *b*-axis of its crystal lattice. In fact, previous measurements on rubrene single-crystal field-effect transistors indicate hole mobility values up to 40 $cm^2V^{-1}s^{-1}$ [28,29]. Furthermore, large arrays of high-performance rubrene single-crystal field-effect transistors with mobilities as high as 2.4 $cm^2/V^{-1}s^{-1}$ and large on/off ratios on the order of $10^7$, have been fabricated on flexible substrates [30]. For these reasons, rubrene is one of the most relevant technological candidates for organic electronics. The superior electronic properties of rubrene, relative to other molecular semiconductors, are mainly attributed to the formation of a dispersive valence band along the *b\** reciprocal lattice vector, shown in Figure 1c. Indeed, density-functional theory calculations indicate a highly dispersive valence band with a bandwidth of 0.4 eV along the Γ-Y path in the Brillouin zone [31,32]. This implies a considerable wavefunction overlap among adjacent rubrene molecules along *b*-axis in real space. In line with this, previous angle-resolved photoemission investigations have observed the valence band dispersion of rubrene [33–36]. Consequently, it is expected that charge transport in rubrene single crystal is essentially mediated by delocalized holes [37]. The current challenge is to control the electronic bands of rubrene through doping without disrupting the single-crystal structure and thus impeding carrier transport.

For this purpose, surface molecular doping can be a more effective strategy than bulk doping. As the electronic properties of organic single crystals ultimately depend on their crystal structure, it is anticipated that bulk doping might have a negative impact. In fact, Hall measurements in bulk-doped rubrene single crystals reported previously by Ohashi *et al.* show decreased Hall mobility. This was attributed to the lattice defects generated by doping that can act as scattering sites, thus reducing the mean free path of the mobile holes [38]. In contrast, introducing a uniform layer of dopants to the surface of rubrene single crystals should not be destructive and could, in principle, alter the near-surface intrinsic electronic character to p-type or n-type.

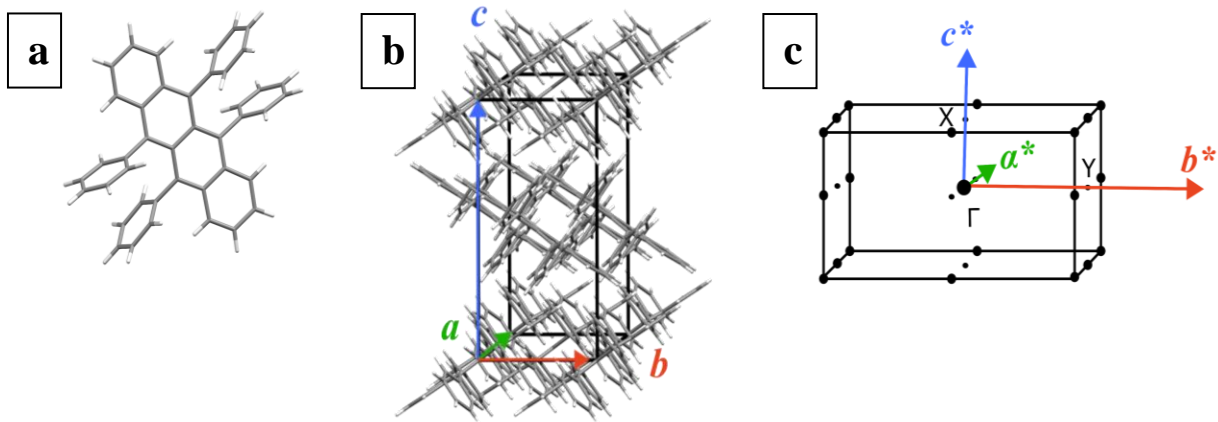

**Figure 1** Molecular structure of rubrene (left). Primitive unit cell of the orthorhombic rubrene single crystal (Cmca) (middle). The first Brillouin zone of the reciprocal lattice of rubrene (right). As indicated by the color, each direct lattice vector corresponds to a reciprocal lattice vector. Each direct lattice vector in orthorhombic crystals is parallel to its corresponding reciprocal lattice vector. In the reciprocal space, a dispersive valence band appears along the ***b*** direction due to the strong wavefunction overlap between the HOMO orbitals along the ***b*** crystallographic direction in real space. The Γ, Y and X correspond to high symmetry points of the Brillouin zone.

In this work, we deposited thin layers of two bulky molecular dopants on rubrene single crystals by thermal evaporation in ultrahigh vacuum. One dopant, Mo(tfd-CO$_2$Me)$_3$, whose chemical structure is shown in Figure 2a, has a high electron affinity of 5.0 eV, measured previously by inverse photoemission spectroscopy [39]. As shown schematically in the energy level diagram of Figure 2c, because the solid-state ionization energy of rubrene almost lies at the same energy, electron transfer is expected to occur from the valence band of rubrene to the lowest unoccupied molecular orbital (LUMO) level of Mo(tfd-CO$_2$Me)$_3$. The electron depletion at the surface of rubrene should push the Fermi level towards the valence band, giving the surface a p-type character. Conversely, when an electron-donating molecule is used, the opposite is anticipated. As an electron donor we used cobaltocene (CoCp$_2$), whose chemical structure is shown in Figure 2b. Cobaltocene has a low solid-state ionization energy of 4.1 eV, reported previously [40]. This can trigger an electron transfer from the singly occupied molecular orbital (SOMO) level of CoCp$_2$ to the conduction band of rubrene. The higher electron concentration at the surface of rubrene would move its Fermi level towards the conduction band, resulting in an n-type conductivity behavior.

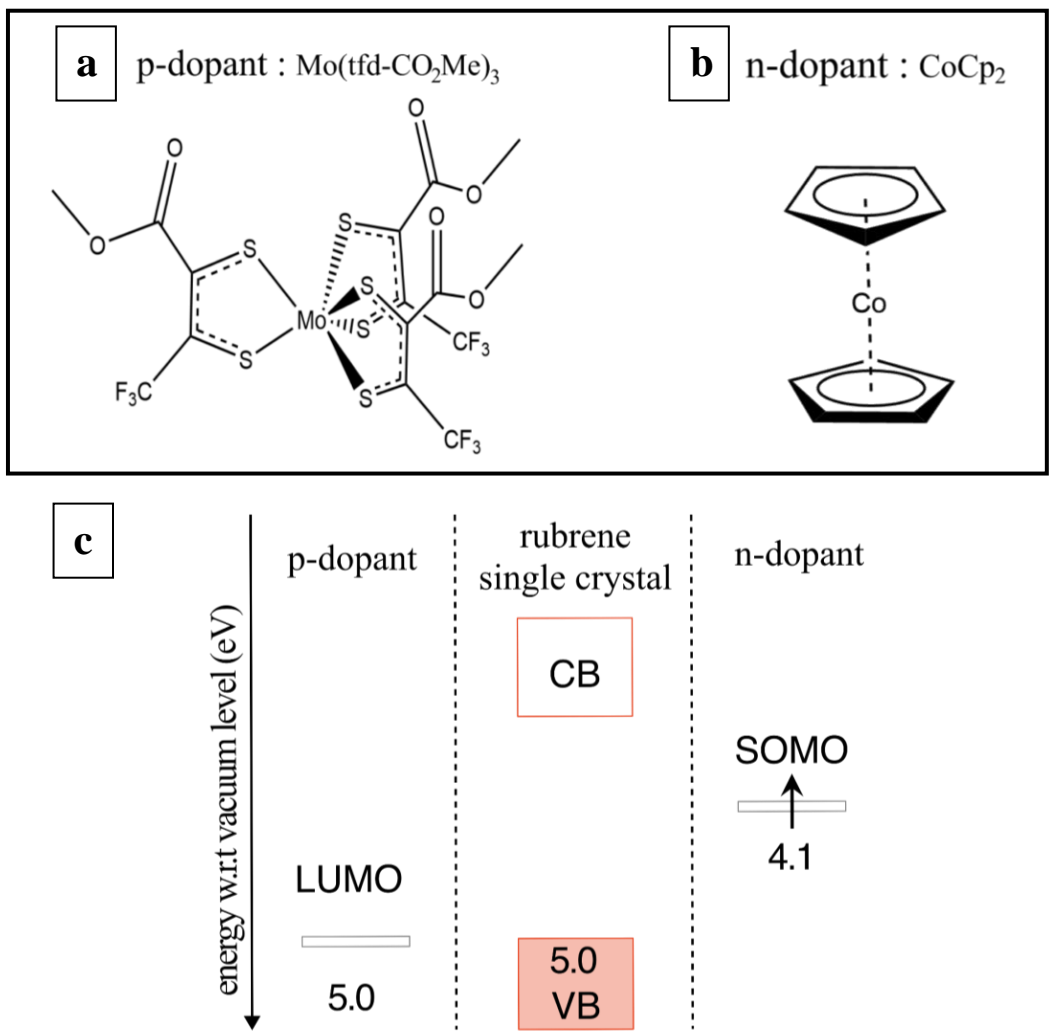

**Figure 2** Chemical structures of Mo(tfd-CO$_2$Me)$_3$ (a) and CoCp$_2$ (b) molecules. The electronic energy levels for the LUMO and SOMO of the p-dopant and n-dopant, respectively, are shown with respect to the vacuum level (c). In the case rubrene single crystal wider boxes are used to represent the valence (VB) and conduction bands (CB). The valence band maximum is measured in this work and found to be 5.0 eV. The energy of the conduction band minimum is tentatively positioned approximately 2.8 eV lower, taking into account the band gap previously measured in rubrene thin films by inverse photoemission spectroscopy [57]. This placement provides only a rough representation as it applies data from thin films to our current study on single crystals.

To test the hypotheses, we performed angle-resolved photoemission spectroscopy (ARPES) measurements on bare and surface-doped rubrene single crystals with the above-mentioned molecular electron donors and acceptors. ARPES allows direct observation of the electronic bands, and we could directly disclose the effect of surface doping on the valence band of rubrene. Our results underpin this non-destructive doping approach, in which the dopants are adsorbed on the surface of rubrene without disrupting its single-crystalline structure.

**Experiment**

Rubrene single crystals were grown via horizontal physical vapor transport [41,42]. For this purpose, a steep temperature gradient is applied across a fused silica glass tube which is located in a furnace. Simultaneously, a 30 sccm $N_2$ (6 N purity) inert gas flow is applied. On the hot side of the furnace, the starting material, ca. 105 mg of purified rubrene powder, is placed and sublimed at 335 °C over 48 h. The sublimed material is transported along the temperature gradient to the cold side, where it recrystallizes, yielding plate-like single crystals with a lateral extension of up to 5 mm. The as-grown crystals were slowly cooled down over 8 h to minimize thermal stress.

Prior to the photoemission experiments, the rubrene single crystals were attached to copper substrates using silver paste to allow sample grounding. Photoemission experiments were performed at the LowDosePES end-station of the BESSY II PM4 beamline of the Helmholtz Zentrum Berlin für Materialien und Energie GmbH (HZB). PM4 is equipped with an angle-resolved time-of-flight (ArTOF) spectrometer that collects the emitted electrons at a broad solid angle of up to 30°. The pulsed excitation source required by the ArTOF analyzers is suitable for investigating radiation sensitive samples such as organics [43]. All experiments were conducted at ambient temperature and in ultrahigh vacuum ($10^{-9}$ mbar). The excitation energy was set to 35 eV corresponding to a short photoelectron mean free path (5-10 Å), for which only electrons from rubrene's uppermost layers contribute to the signal. The experimental setup provides a 20 meV energy resolution and a 0.09° angular resolution.

We measured two rubrene single-crystal samples, the electronic band structures of which were essentially identical (Figure S1, supporting information). A minor discrepancy of approximately 50 meV in the binding energy position of the valence peaks between the two samples can be observed by the results of the quantitative analysis on the valence peaks (Table S1, supporting information). This discrepancy could potentially be attributed to the inherent uncertainty associated with the measurements of samples of varying thicknesses, or to minute variations in crystal growth conditions within the same or different production batches. During the measurement, consecutive energy-distribution curves (EDCs) recorded over equal time durations revealed the impact of sample charging: broadening and shifting of the HOMO peak towards higher binding energy. In photoemission from materials having low conductivity, such as organics, the buildup of a positive surface charge due to the removal of electrons is typically a hurdle for the accurate interpretation of photoemission spectra. But it can usually be compensated by taking advantage of their photoconductivity induced by an external source of light [44]. In our case, we used a continuous wave laser of 473 nm wavelength to constantly illuminate the sample. To neutralize the radiation induced charging, a nominal laser power density of 0.3 mW/cm$^2$ was required. (Figure S2, supporting information).

Surface doping was achieved in-situ by thermally evaporating the molecular dopants on bare rubrene single crystals. p-Doping was achieved by depositing a thin layer of Mo(tfd-CO$_2$Me)$_3$ at a constant evaporation rate of 0.1 Å/s, as measured by a quartz crystal microbalance. The final nominal thickness of the molecular layer was 1.5 nm. In the case of the n-doping with CoCp$_2$ the film thickness could not be determined due to the high volatility of CoCp$_2$. Instead, we observed Instead, we monitored the changes in rubrene's valence band structure throughout three successive evaporations, each of an increasing duration.

## Results and Discussion

I. Band structure of the bare rubrene single crystal

In Figures 3a-3d, we show the angle-resolved EDCs and the corresponding 2D spectra of bare rubrene along the Γ-Y and Γ-X directions of the Brillouin zone, respectively. Along the Γ-X direction, Figures 3b and 3d, there is no clear indication of dispersive features in the valence region. In the Γ-Y direction, Figures 3a and 3c, there is a visible dispersive peak (H$_1$), which can be associated with the HOMO derived valence band of rubrene single crystals, as demonstrated in earlier experimental and theoretical studies [31–34]. The maximum of the valence band is at the Γ point, 0.64 eV below the Fermi level, determined by the peak maximum of $H_1$ at the $\Gamma$ point. Notably, the $\Gamma$-Y direction shows two additional features, A and B, which appear at fixed binding energy positions below the band maximum and have also been observed in prior ARPES measurements [33,34,45]. Previous studies, both theoretical and experimental, demonstrate the occurrence of a second dispersive band along the $\Gamma$-Y direction due to the presence of two inequivalent molecules in the unit cell [27,32,33]. This can explain the A feature below $H_1$. A peak analysis of the EDC at the $\Gamma$ point revealed that the energy separation between $H_1$ and A is roughly 0.3 eV (Figure S3 and Table S1, supporting information). The observed energy separation does not fully match the theoretical value of 0.20 eV, although such a difference is expected when density-functional theory fails to capture the correct many-body correlation effects of the excited system. In fact, it has been demonstrated before for pentacene that molecular vibration and the presence of disorder can explain the greater separation between the two HOMO-derived bands [46].

Additionally, the origin of feature B, unexplained by previous density-functional theory calculations of the band structure of rubrene, may be attributed to scattering processes during photoemission. These could be a result of both thermal disorder and structural disorder on the surface, possibly induced by the prior exposure of the crystals to ambient conditions [47]. For example, scattering by vibrations and impurities can spread the discrete electron states over a range of energies and momenta giving rise to an incoherent background component to the photoemission spectrum [48]. According to Ciuchi *et al.*, photoelectrons can be scattered at room temperature by the intramolecular vibrations of rubrene, resulting in broadened spectral linewidths [49]. The B feature, appearing as a k-independent uniform background contributing to the broadening of the $H_1$ peak, can be considered the incoherent component of the total *k*-resolved spectral intensity. Previous photoemission studies have demonstrated that, in the presence of dynamic disorder, the total *k*-resolved photoelectron spectrum ($I_{tot}(k,E)$) can be described as the sum of a coherent spectrum associated with direct electronic transitions ($I_{coh}(k,E)$), and an incoherent background ($I_{incoh}(k,E)$):

$$I_{tot}(k,E) = W \cdot I_{coh}(k,E) + I_{incoh}(k,E) \qquad (1)$$

Here, $W$ represents the Debye-Waller factor [50,51] [49,50]. Despite not knowing the Debye-Waller factor it is still possible to approximate the coherent part of the spectrum associated with rubrene's intrinsic valence band by subtracting a reasonable incoherent background.

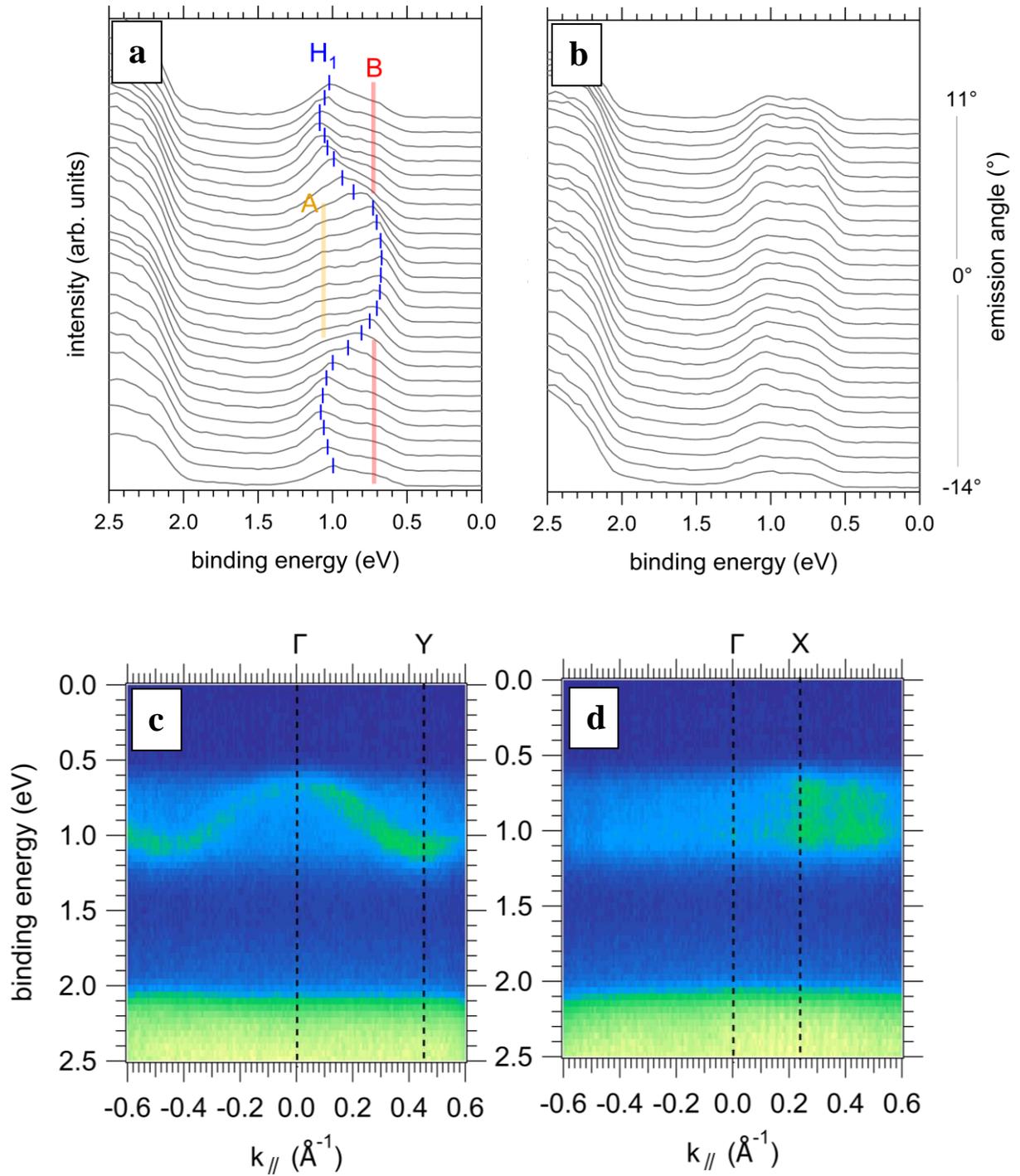

**Figure 3** Angle-resolved EDCs along Γ-Y (a), and Γ-X (b) directions of the Brillouin zone. The short blue lines indicate the peak position of the dispersive H1 peak and the yellow and red lines show the non-dispersive A and B satellite features of the spectrum. The respective 2D spectra of the Γ-Y (c) and Γ-X (d) show valence band of rubrene. The black dashed lines indicate the positions of the Γ, Y and X high symmetry points which are also the boundaries of the Brillouin zone (BZ).

In such cases, the incoherent background resembles the *k*-integrated density of states [51,52]. Thus, subtracting the experimental *k*-integrated EDC as a background from every *k*-resolved EDC simplifies the quantitative analysis for determining the intrinsic band parameters (Figures S4 and S5 supporting information). The band parameters can be determined by approximating the HOMO-derived valence band with the energy dispersion relation of the one-dimensional tight-binding model:

$$E_B(k) = E_C + 2t \cdot \cos(ak) \qquad (2)$$

where $E_B$ is the electron's binding energy, $k$ the electron's momentum, $E_C$ the energy of the center of the valence band, $t$ the transfer integral, and $a$ the lattice constant. Moreover, the hole effective mass at the $\Gamma$ point, $m_{TB}^*$, is given by:

$$m_{TB}^* = -\frac{\hbar^2}{2ta^2} \qquad (3)$$

The above relations are used for the analysis of band structure properties further below.

### II. Surface molecular doping

The 2D intensity maps of bare and surface-doped rubrene are depicted in Figures 4a-4d. In the case of Mo(tfd-CO$_2$Me)$_3$, the deposited layer of molecular acceptors shifted the valence band of rubrene 90 meV toward the Fermi level, as seen in Figures 4a and 4b. In contrast, the molecular donor CoCp$_2$ induced a 140 meV shift of the valence band in the opposite direction, as shown in Figures 4c and 4d. In either case, the shift of the valence band can be attributed to electron-transfer reactions at the interface between the dopants and rubrene. In particular, the deposited layer of molecular acceptors triggers an electron transfer from the highest occupied valence states of rubrene close to the interface to the LUMO of Mo(tfd-CO$_2$Me)$_3$ in order to achieve electronic equilibrium. The redistribution of charges (i.e., space-charge accumulation) generates an electric field close to the interface, which bends the valence band upwards and is associated with an increased concentration of holes at the surface of rubrene. Conversely, the deposition of donor molecules causes an electron transfer from the SOMO of CoCp$_2$ to the lowest unoccupied states of rubrene. The increased concentration of electrons at the surface of rubrene induced a downward bending of the valence band. In addition, we observed that the valence band along the $\Gamma$-X direction exhibited similar shifts due to surface doping (as shown in Figure S6 of the supporting information). This observation suggests that the doping process likely induced a uniform charge distribution on the surface, leading to a concurrent shift in all electronic states.

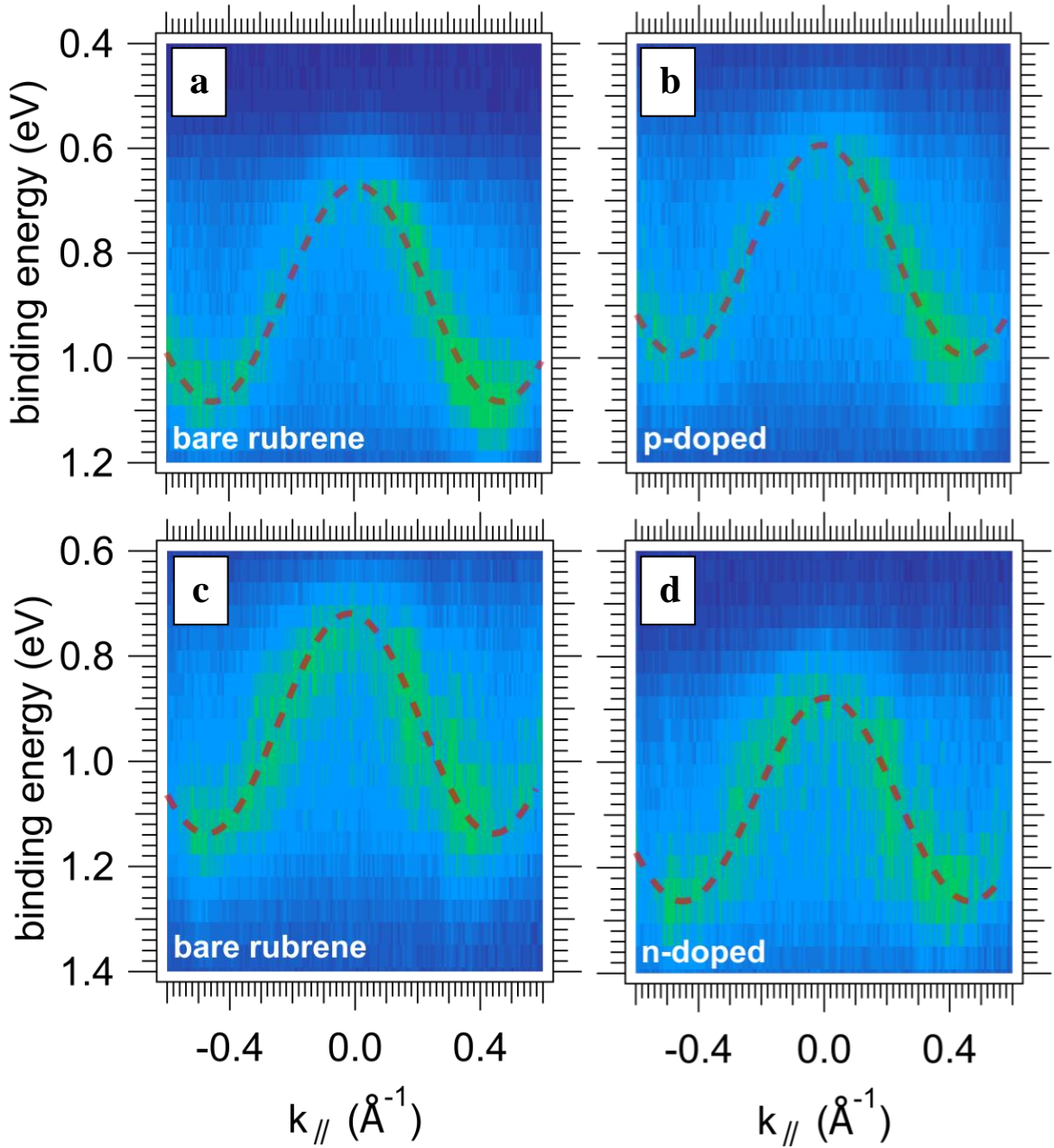

**Figure 4** 2D spectra of the bare (undoped) rubrene (a), (c), surface p-doped rubrene (b) and surface n-doped rubrene. The red curve represents the least-square fit of the valence band approximated by the 1D tight binding dispersion relation.

Assuming that the induced charge carriers at the band edges behave as free particles and that they obey Boltzmann statistics we can roughly estimate the induced concentration of holes by the following expression:

$$\frac{p_f}{p_i} = \exp\left[-\frac{\Delta E_V}{k_b T}\right] \quad (4)$$

where $p_i$ is the initial concentration of holes for the undoped rubrene, $p_f$ the final concentration of holes after surface doping and $\Delta E_V$ the energy shift of the valence band [53]. Given that at room temperature the thermal energy is $k_b T = 25$ meV, then the valence-band energy shifts of -90 meV and 140 meV could effectively modulate the hole concentration at the surface by up to two orders of magnitude. Most importantly, it appears that the surface molecular doping had no apparent effect on the shape of the rubrene valence band. A quantitative analysis, the results of which are listed in Table 1, revealed that the band parameters, transfer integral, and hole effective mass, remain essentially constant. Therefore, it is reasonable to assume that the deposition of Mo(tfd-CO$_2$Me)$_3$ and CoCp$_2$, and the induced charge concentrations, do not result in structural deformation of the surface lattice of rubrene nor in anisotropic charge-carrier distribution along certain directions.

**Table 1** Band parameters of 1D-tight binding approximation where ($E_C$) the center of the valence band, ($\Delta E_C$) the shift of the valence band center after doping, ($a$) lattice parameter corresponding to $b$ crystallographic axis of rubrene single crystal, ($t$) transfer integral and ($m^*_{TB}$) hole effective mass.

|  | $E_C$ (eV) | $\Delta E_C$ (meV) | $a$ (Å) | $t$ (meV) | $|m^*_{TB}|$ ($m_0$) |
|---|---|---|---|---|---|
| **bare rubrene** | 0.88 ± 0.01 | - | 6.8 ± 0.1 | 103 ± 10 | 0.8 ± 0.1 |
| **p-doped** | 0.79 ± 0.01 | -90 ± 10 | 6.8 ± 0.1 | 100 ± 10 | 0.8 ± 0.1 |
| **bare rubrene** | 0.93 ± 0.01 | - | 6.9 ± 0.1 | 105 ± 10 | 0.7 ± 0.1 |
| **n-doped** | 1.07 ± 0.01 | 140 ± 10 | 6.9 ± 0.1 | 101 ± 10 | 0.8 ± 0.1 |

The magnitude of the valence band shifts, related to doping efficiency, is noteworthy, as it does not appear to be strictly determined by the energy offsets between the frontier energy levels of rubrene and the dopants. One might expect larger changes, even for such small energy differences.

Several distinct and interrelated factors may explain this situation. First, the small shift in the case of the p-dopant could be related to the low dopant concentration on the surface, as indicated by the small signals observed in the X-ray photoemission spectra (Figure S7, supporting information). Second, materials inherently contain native defects, the origin, concentration, and spatial distribution of which largely depend on the crystal growth conditions and the post-treatment of the crystals after growth. These native defects can generate an intra-bandgap density of states, acting as electron donors or acceptors, and consequently, compensating for further doping of the surface. In this scenario, the Fermi level becomes pinned by the defect states, causing the valence band to be fixed at a certain position, irrespective of the dopant concentrations [54,55]. Lastly, intermolecular interactions at the interface can significantly modify the energetics of the donating and accepting energy levels, leading to unfavorable energy barriers that impede charge-carrier hopping across the interface [56].

## Conclusions

In conclusion, our photoemission results demonstrate that surface doping by thermal evaporation of molecular thin layers can indeed change the intrinsic electronic character of rubrene single crystal surfaces to more n-type or p-type. This is attributed to the observed shifts of the rubrene valence band relative to the Fermi level. Importantly, the doping is accomplished without disrupting the surface structure of single-crystal rubrene, a prerequisite to preserving the HOMO-derived valence band. A next step could be charge-transport measurements to test how the surface molecular doping of rubrene single crystals affects device characteristics. Furthermore, in order to comprehend the magnitude of the observed shifts and identify specific routes for the efficient doping of organic semiconductors it is necessary to fully understand the reasons that may limit the surface molecular doping of rubrene. This could involve the use of different molecular dopants and a systematic investigation of both the microscopic interactions between the semiconductor host and the dopants as well as the potential thermodynamic limitations of doping organic semiconductors containing native defects. These points should form the basis of our future studies.

## Acknowledgments

We gratefully acknowledge the beamline scientists of PM4 end-station Dr. Erika Giangrisostomi and Dr. Ruslan Ovsyannicov for their kind support in the course of the ArTOF experiments. We also thank Dr. Patrick Amsalem and Dr. Ross Warren for the insightful discussions. This project has received funding from the DFG project 239543752 and the European Union's Horizon 2020 research and innovation programme under the Marie Skłodowska-Curie grant agreement No 811284 (UHMOB). S.H. and J.P. acknowledge financial support by the Bavarian State Ministry of Science, Research, and the Arts (Collaborative Research Network 'Solar Technologies Go Hybrid'). Y.Z, S.B., and S.R.M. thank the National Science Foundation for support through the DMREF program (DMR-1729737).

# Supporting information

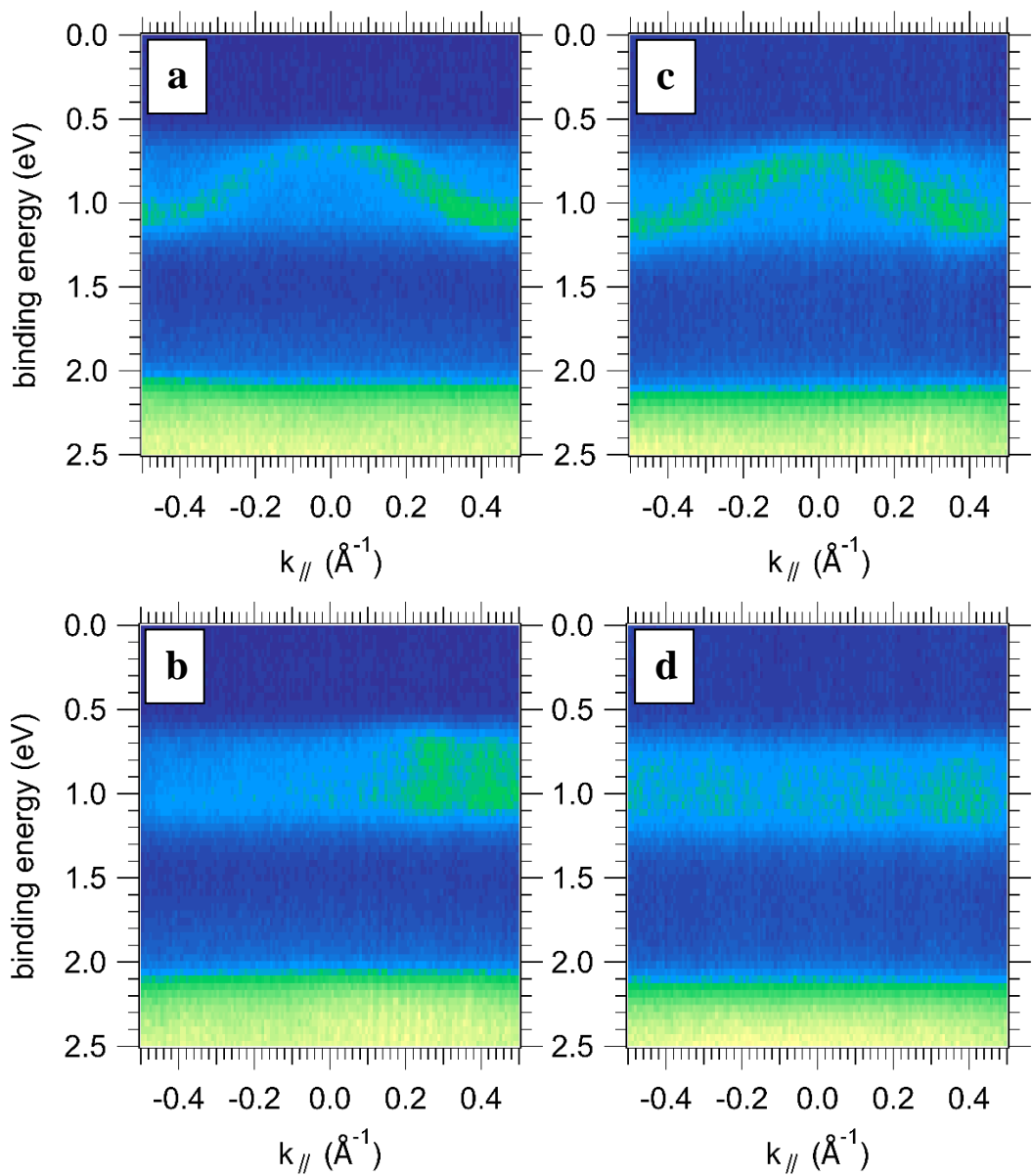

**Supplementary Figure S1: 2D spectra of the two bare rubrene single crystal samples measured prior to the surface doping.** (a), (b) 2D spectra of the first single crystal rubrene sample along the Γ-Y and Γ-X directions, respectively. (c), (d) 2D spectra of the second single crystal rubrene sample along the Γ-Y and Γ-X directions, respectively. The 2D spectra of the two single crystal rubrene samples show strong qualitative agreement.

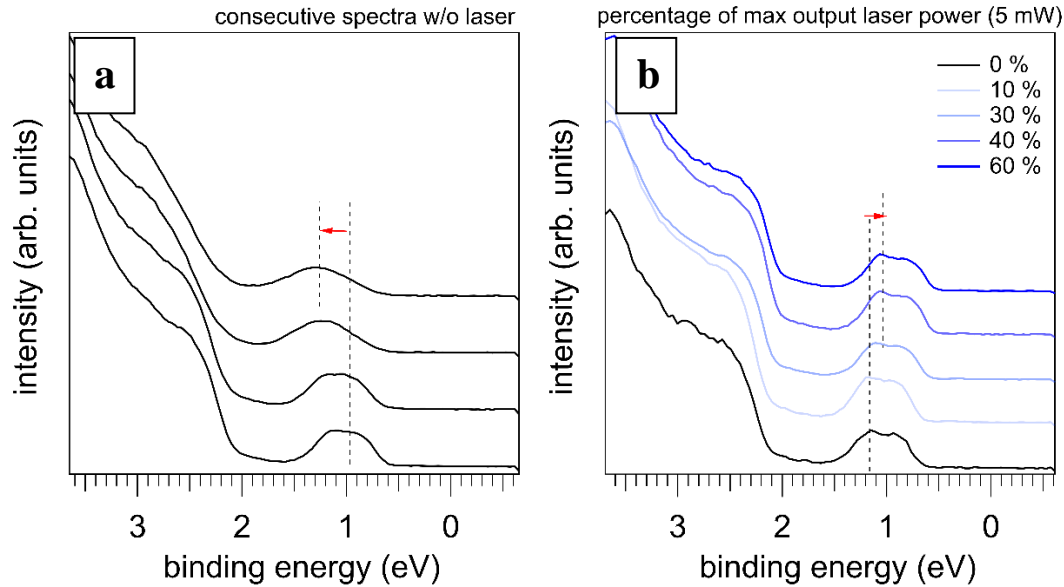

**Supplementary Figure S2: Influence of surface charging and its compensation using an external source of light (continuous wave blue laser of 473 nm and with 5 mW max output power).** (a) In the absence of laser, consecutive EDCs of equal time durations revealed that the HOMO peak became broader and drifted towards higher binding energies, indicating strong effects of surface charging. By illuminating the sample with a blue laser we could compensate the effect of charging (b). The progressive increase of the laser output power shifts the HOMO peak of rubrene back to lower binding energies. With laser power greater than 40 percent, the position and the shape of the peak stabilized. Given that the diameter of the defocused laser beam is approximately 16 mm, 60 percent of laser power corresponds to a nominal power density of 0.3 mW/cm$^2$.

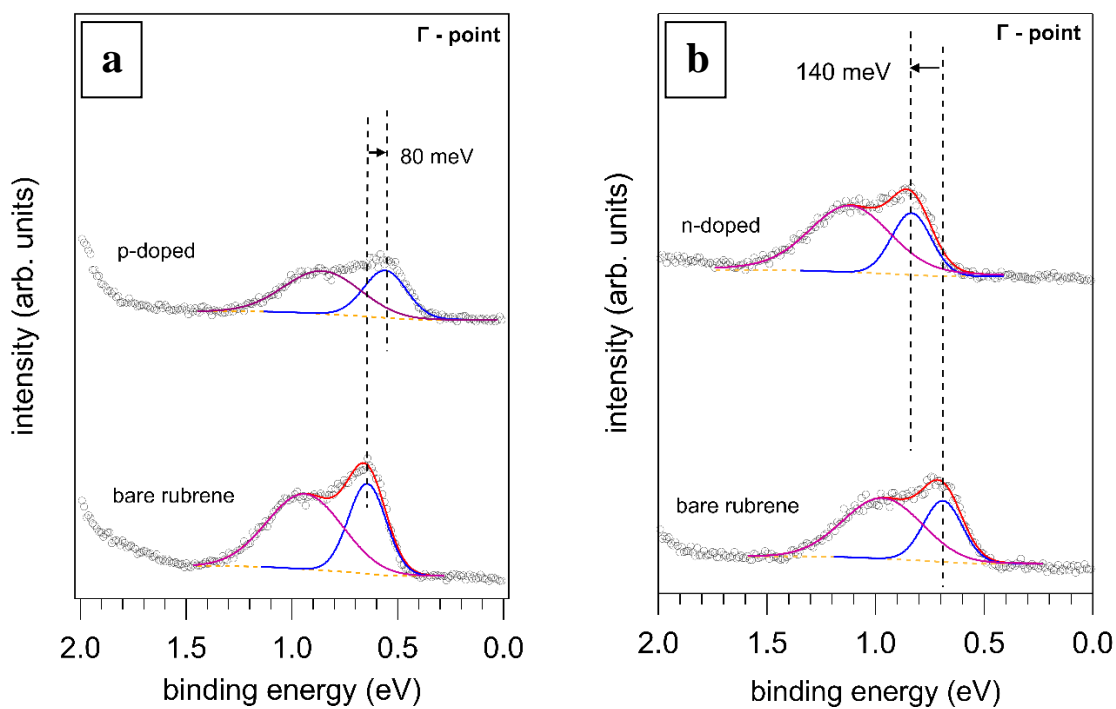

**Supplementary Figure S3: Quantitative analysis of the HOMO peak at the Γ point of bare and surface-doped rubrene.** The total HOMO peak was deconvoluted using two Voigt functions to account for the dispersive $H_1$ peak and the satellite feature A. Voigt functions are preferred for fitting photoelectron peaks, as they are convolutions of Gaussian and Lorentzian profiles that represent the instrumental broadening and the natural broadening due to the lifetime of the photoexcited state, respectively. The empty circles indicate the raw data points, the blue Voigt function represents the $H_1$ peak, whereas the purple is associated to the A satellite peak. The yellow dashed background corresponds to the Shirley background. The red curves represent the total least-squares fitting curve of the valence peak. The black vertical dashed lines depict the positions of the $H_1$ peaks before and after doping. The fitting parameters are listed in Table S1. For $Mo(tfd-CO_2Me)_3$ (p-doping), the $H_1$ peak shows a shift of 80 meV towards lower binding energies (a). In the case of $CoCp_2$ (n-doping), the $H_1$ peak shifts by 140 meV towards higher binding energies (b). The intensity from the HOMO peak of rubrene is attenuated due to the deposition of the molecular dopant layers on the surface.

**Supplementary Table S1: Least-square fitting parameters of Figure S4.** $E_{H1}$ is the energy position of the dispersive $H_1$ peak, $E_A$ the energy position of the satellite A feature, **FWHM**$_{H1}$ and **FWHM**$_A$ the full width at half maximum of the peak $H_1$ and A, respectively.

|  | $E_{H1}$ (eV) | FWHM$_{H1}$ (eV) | $E_A$ (eV) | FWHM$_A$ (eV) |
|---|---|---|---|---|
| **bare rubrene** | 0.64 ± 0.01 | 0.21 ± 0.04 | 0.94 ± 0.02 | 0.42 ± 0.08 |
| **p-doped** | 0.56 ± 0.01 | 0.24 ± 0.05 | 0.86 ± 0.02 | 0.42 ± 0.09 |
| **bare rubrene** | 0.69 ± 0.01 | 0.21 ± 0.05 | 0.97 ± 0.02 | 0.43 ± 0.08 |
| **n-doped** | 0.83 ± 0.01 | 0.21 ± 0.05 | 1.12 ± 0.02 | 0.44 ± 0.08 |

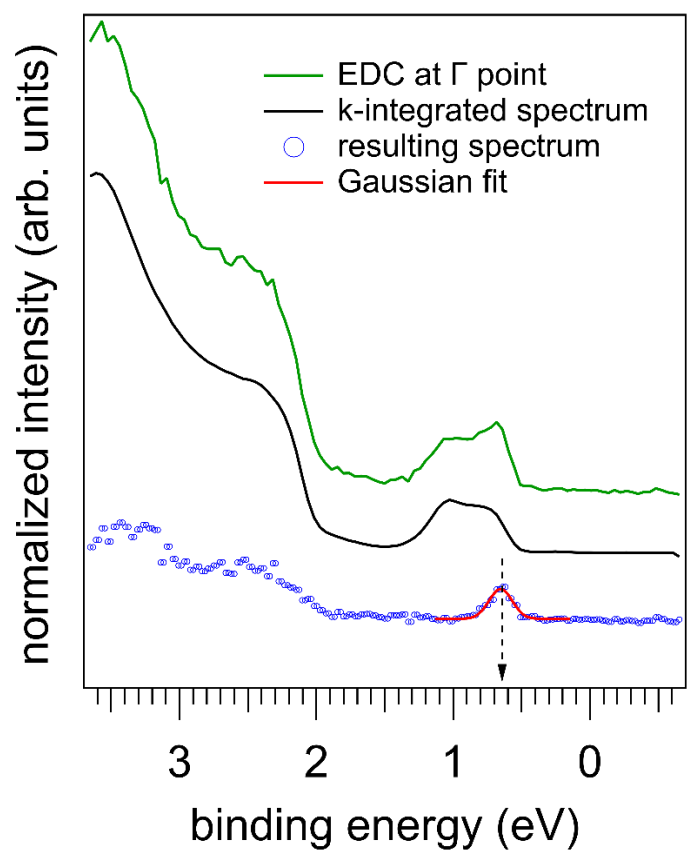

**Supplementary Figure S4: Background subtraction procedure applied to the EDC at the Γ point.** The exact methodology for subtracting the background from the 2D spectra can be clearly depicted in the figure above. The green curve represents the k-resolved EDC, the black curve corresponds to the k-integrated spectrum that has been normalized relative to the satellite feature's intensity. The blue dotted line represents the spectrum remaining after subtraction of the k-resolved EDC and the k-integrated spectrum. The red line corresponds to the Gaussian fit of the residual spectrum, from which the energy position of the dispersive $H_1$ peak is determined.

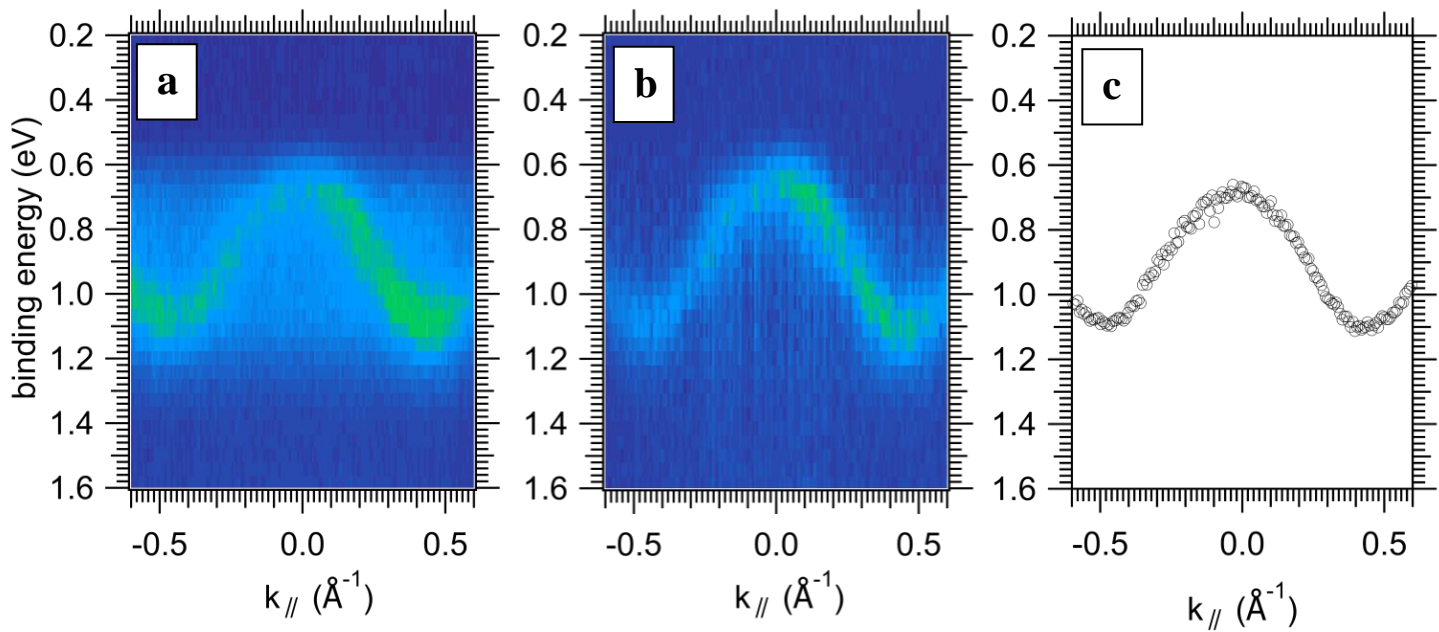

**Supplementary Figure S5: Background subtraction applied to the 2D spectra.** By using the background subtraction approach illustrated in Figure S5 to all k-resolved EDCs, we were able to eliminate the contribution of the A and B satellites attributed to scattering processes due to the dynamic disorder at room temperature. Specifically, (a) shows the total 2D spectrum of bare rubrene and (b) displays the spectrum remaining after background subtraction. Using a batch fitting script implemented in Igor Pro 9, each residual k-resolved profile, was fitted by one Gaussian peak to determine the energy position of the $H_1$ peak. (c) shows the extracted (E,k) data points corresponding to the $H_1$ dispersive peak.

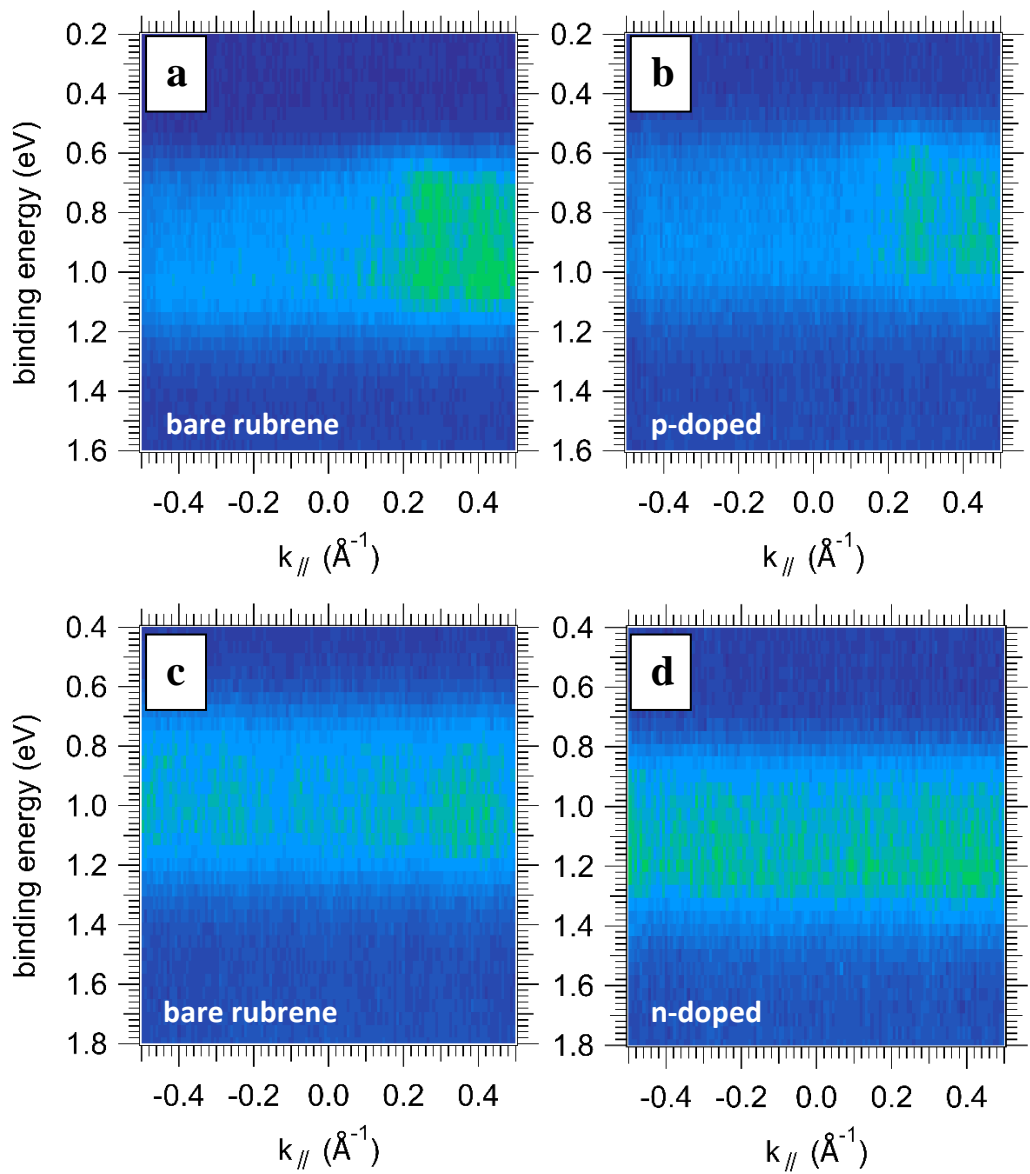

**Supplementary Figure S6: Band bending of the valence band of rubrene along the Γ-X directions due to the surface molecular doping.** As mentioned in the main text for the case of the valence band along the Γ-Y direction, the charge redistribution at the interface induced by doping results in the bending of all electronic states, including the valence band along the Γ-X direction. In particular, p-doping with Mo(tfd-CO$_2$Me)$_3$ shifts the valence band towards the Fermi level, (a) to (b). Similarly, when rubrene is n-doped with CoCp$_2$ the valence band shifts away from the Fermi level (c) to (d).

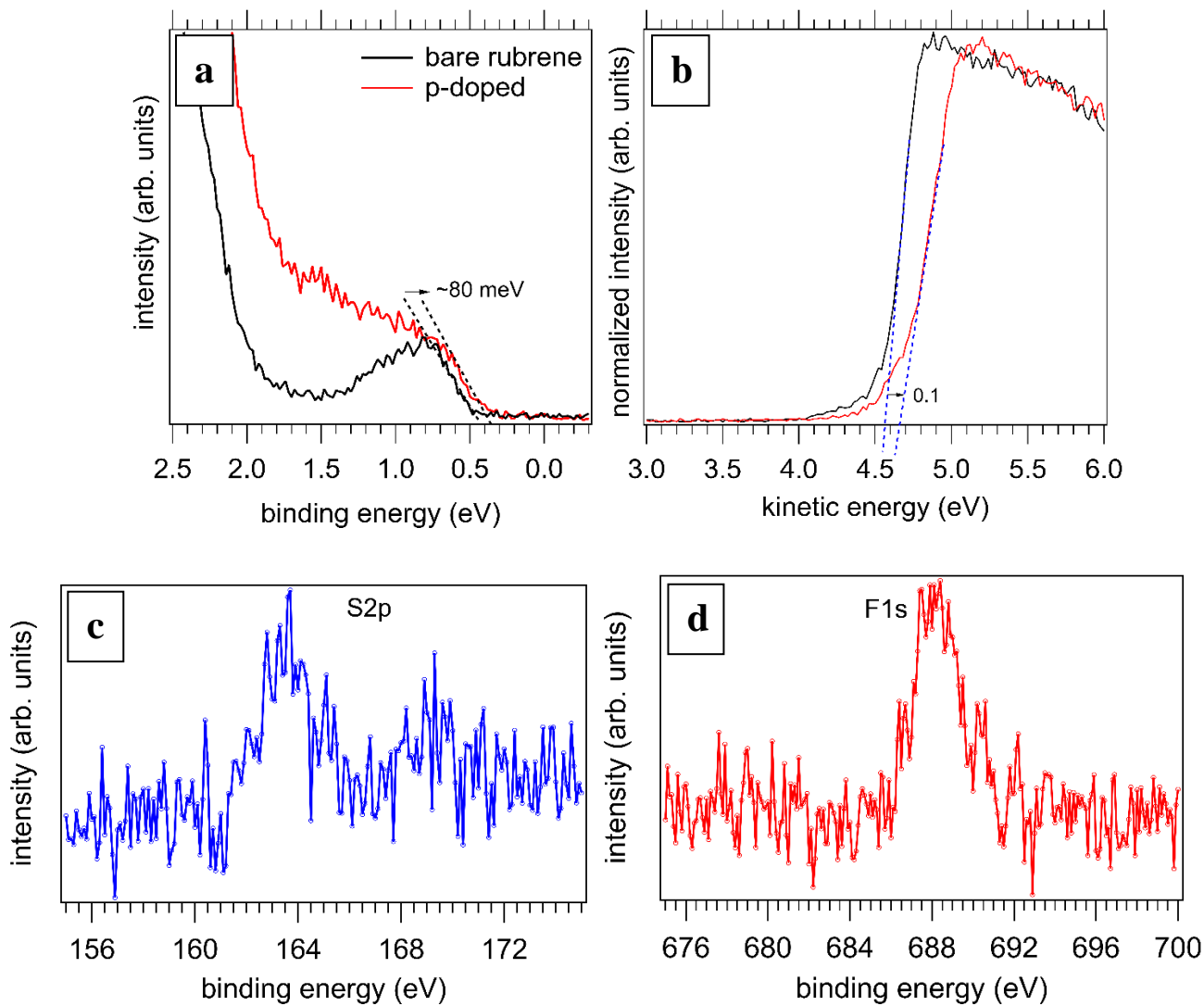

**Supplementary Figure S7: Additional ultraviolet (UPS) and X-ray (XPS) photoemission spectra.** This figure provides further support to the surface doping of bare rubrene with the molecular acceptor Mo(tfd-CO$_2$Me)$_3$. Additional UPS and XPS measurements were performed on a standard setup with a He discharge lamp (He-I 21.22 eV) and a hemispherical analyzer (Specs, PHOIBOS 100). The valence region of rubrene (a), was obtained at the Γ point before and after the evaporation of 1.5 nm Mo(tfd-CO$_2$Me)$_3$. In agreement to what we found in our ArTOF measurements, the evaporation of the dopants induced a shift of the valence peak by approximately 80 meV, as determined by a linear extrapolation of the HOMO peak onsets. After doping the spectrum becomes less structured with the valence peak hidden in the background intensity. This is expected as the dopant adlayer can prevent the electrons from the covered rubrene's surface from being phootemitted. The UPS valence region was recorded with pass energy 10 eV that allows energy resolution of 120 meV. (b) shows the secondary electron cut-off region (SECO). The SECO was obtained using -10 V bias voltage and pass energy 2 eV allowing an energy resolution of 60 meV. The work function can be determined by a linear extrapolation of the onset of the SECO. P-doping increases the work function by roughly 0.1 eV. The shift of the work function is consistent with the band bending of the valence band (80 meV). (c) and (d) show the XPS signals of Sulfur's 2p and Fluorine's 1s core levels, which are present in Mo(tfd-CO$_2$Me)$_3$.

28